\documentclass[%
twocolumn,
superscriptaddress,
 amsmath,amssymb,
 prc
 aps,
]{revtex4-1}

\def\lmax{$\mathrm{L}_{\mathrm{lab}}^{\mathrm{max}}$}

\usepackage{romanbar}
\usepackage{graphicx}
\usepackage{dcolumn}
\usepackage{bm}
\usepackage{SIunits}
\usepackage{array}

\begin{document}

\title{The First Direct Search for Inelastic Boosted Dark Matter with COSINE-100}

\author{C.~Ha}
\affiliation{Center for Underground Physics, Institute for Basic Science (IBS), Daejeon 34126, Republic of Korea}
\author{G.~Adhikari}
\affiliation{Department of Physics, Sejong University, Seoul 05006, Republic of Korea}
\author{P.~Adhikari}
\affiliation{Department of Physics, Sejong University, Seoul 05006, Republic of Korea}
\author{E.~Barbosa~de~Souza}
\affiliation{Department of Physics and Wright Laboratory, Yale University, New Haven, CT 06520, USA}
\author{N.~Carlin}
\affiliation{Physics Institute, University of S\~{a}o Paulo, 05508-090, S\~{a}o Paulo, Brazil}
\author{S.~Choi}
\affiliation{Department of Physics and Astronomy, Seoul National University, Seoul 08826, Republic of Korea} 
\author{M.~Djamal}
\affiliation{Department of Physics, Bandung Institute of Technology, Bandung 40132, Indonesia}
\author{A.~C.~Ezeribe}
\affiliation{Department of Physics and Astronomy, University of Sheffield, Sheffield S3 7RH, United Kingdom}
\author{I.~S.~Hahn}
\affiliation{Department of Science Education, Ewha Womans University, Seoul 03760, Republic of Korea} 
\author{E.~J.~Jeon}
\affiliation{Center for Underground Physics, Institute for Basic Science (IBS), Daejeon 34126, Republic of Korea}
\author{J.~H.~Jo}
\affiliation{Department of Physics and Wright Laboratory, Yale University, New Haven, CT 06520, USA}
\author{H.~W.~Joo}
\affiliation{Department of Physics and Astronomy, Seoul National University, Seoul 08826, Republic of Korea}
\author{W.~G.~Kang}
\affiliation{Center for Underground Physics, Institute for Basic Science (IBS), Daejeon 34126, Republic of Korea}
\author{W.~Kang}
\affiliation{Department of Physics, Sungkyunkwan University, Suwon 16419, Republic of Korea}
\author{M.~Kauer}
\affiliation{Department of Physics and Wisconsin IceCube Particle Astrophysics Center, University of Wisconsin-Madison, Madison, WI 53706, USA}
\author{G.~S.~Kim}
\affiliation{Department of Physics, Kyungpook National University, Daegu 41566, Republic of Korea}
\author{H.~Kim}
\affiliation{Center for Underground Physics, Institute for Basic Science (IBS), Daejeon 34126, Republic of Korea}
\author{H.~J.~Kim}
\affiliation{Department of Physics, Kyungpook National University, Daegu 41566, Republic of Korea}
\author{K.~W.~Kim}
\affiliation{Center for Underground Physics, Institute for Basic Science (IBS), Daejeon 34126, Republic of Korea}
\author{N.~Y.~Kim}
\affiliation{Center for Underground Physics, Institute for Basic Science (IBS), Daejeon 34126, Republic of Korea}
\author{S.~K.~Kim}
\affiliation{Department of Physics and Astronomy, Seoul National University, Seoul 08826, Republic of Korea}
\author{Y.~D.~Kim}
\affiliation{Center for Underground Physics, Institute for Basic Science (IBS), Daejeon 34126, Republic of Korea}
\affiliation{Department of Physics, Sejong University, Seoul 05006, Republic of Korea}
\author{Y.~H.~Kim}
\affiliation{Center for Underground Physics, Institute for Basic Science (IBS), Daejeon 34126, Republic of Korea}
\affiliation{Korea Research Institute of Standards and Science, Daejeon 34113, Republic of Korea}
\author{Y.~J.~Ko}
\affiliation{Center for Underground Physics, Institute for Basic Science (IBS), Daejeon 34126, Republic of Korea}
\author{V.~A.~Kudryavtsev}
\affiliation{Department of Physics and Astronomy, University of Sheffield, Sheffield S3 7RH, United Kingdom}
\author{H.~S.~Lee}
\email{hyunsulee@ibs.re.kr}
\affiliation{Center for Underground Physics, Institute for Basic Science (IBS), Daejeon 34126, Republic of Korea}
\author{J.~Lee}
\affiliation{Center for Underground Physics, Institute for Basic Science (IBS), Daejeon 34126, Republic of Korea}
\author{J.~Y.~Lee}
\affiliation{Department of Physics, Kyungpook National University, Daegu 41566, Republic of Korea}
\author{M.~H.~Lee}
\affiliation{Center for Underground Physics, Institute for Basic Science (IBS), Daejeon 34126, Republic of Korea}
\author{D.~S.~Leonard}
\affiliation{Center for Underground Physics, Institute for Basic Science (IBS), Daejeon 34126, Republic of Korea}
\author{W.~A.~Lynch}
\affiliation{Department of Physics and Astronomy, University of Sheffield, Sheffield S3 7RH, United Kingdom}
\author{R.~H.~Maruyama}
\affiliation{Department of Physics and Wright Laboratory, Yale University, New Haven, CT 06520, USA}
\author{F.~Mouton}
\affiliation{Department of Physics and Astronomy, University of Sheffield, Sheffield S3 7RH, United Kingdom}
\author{S.~L.~Olsen}
\affiliation{Center for Underground Physics, Institute for Basic Science (IBS), Daejeon 34126, Republic of Korea}
\author{B.~J.~Park}
\affiliation{IBS School, University of Science and Technology (UST), Daejeon 34113, Republic of Korea}
\author{H.~K.~Park}
\affiliation{Department of Accelerator Science, Korea University, Sejong 30019, Republic of Korea}
\author{H.~S.~Park}
\affiliation{Korea Research Institute of Standards and Science, Daejeon 34113, Republic of Korea}
\author{K.~S.~Park}
\affiliation{Center for Underground Physics, Institute for Basic Science (IBS), Daejeon 34126, Republic of Korea}
\author{R.~L.~C.~Pitta}
\affiliation{Physics Institute, University of S\~{a}o Paulo, 05508-090, S\~{a}o Paulo, Brazil}
\author{H.~Prihtiadi}
\affiliation{Department of Physics, Bandung Institute of Technology, Bandung 40132, Indonesia}
\author{S.~J.~Ra}
\affiliation{Center for Underground Physics, Institute for Basic Science (IBS), Daejeon 34126, Republic of Korea}
\author{C.~Rott}
\affiliation{Department of Physics, Sungkyunkwan University, Suwon 16419, Republic of Korea}
\author{K.~A.~Shin}
\affiliation{Center for Underground Physics, Institute for Basic Science (IBS), Daejeon 34126, Republic of Korea}
\author{A.~Scarff}
\affiliation{Department of Physics and Astronomy, University of Sheffield, Sheffield S3 7RH, United Kingdom}
\author{N.~J.~C.~Spooner}
\affiliation{Department of Physics and Astronomy, University of Sheffield, Sheffield S3 7RH, United Kingdom}
\author{W.~G.~Thompson}
\affiliation{Department of Physics and Wright Laboratory, Yale University, New Haven, CT 06520, USA}
\author{L.~Yang}
\affiliation{Department of Physics, University of Illinois at Urbana-Champaign, Urbana, IL 61801, USA}
\author{G.~H.~Yu}
\affiliation{Department of Physics, Sungkyunkwan University, Suwon 16419, Republic of Korea}
\collaboration{COSINE-100 Collaboration}
\date{\today}

\begin{abstract}
A search for inelastic boosted dark matter~(iBDM) using the COSINE-100 detector with 59.5\,days of data is presented. 
This relativistic dark matter is theorized to interact with the target material through inelastic scattering with electrons, creating a heavier state that subsequently produces standard model particles, such as an electron-positron pair. 
In this study, we search for this electron-positron pair in coincidence with the initially scattered electron as a signature for an iBDM interaction. No excess over the predicted background event rate is observed. Therefore, we present limits on iBDM interactions under various hypotheses, one of which allows us to explore an area of the dark photon parameter space that has not yet been covered by other experiments. This is the first experimental search for iBDM using a terrestrial detector. 
\end{abstract}
\maketitle

A number of astrophysical observations provide evidence that the dominant matter component of the Universe is not ordinary matter, but rather non-baryonic dark matter~\cite{Clowe:2006eq,Ade:2015xua}. 
A tremendous effort to search for dark matter has been pursued by direct detection experiments~\cite{gaitskell04,Battaglieri:2017aum,Aprile:2018dbl}, indirect detection experiments~\cite{Choi:2015ara,Aartsen:2016zhm,Conrad:2017pms}, and collider experiments~\cite{Sirunyan:2018gka,Aaboud:2017dor,dmcol} with no success~\cite{Tanabashi2018}.  
This motivates searches for new types of dark matter, such as light-mass models~\cite{Essig:2011nj,Essig:2013lka,Hochberg:2016ajh,Knapen:2017xzo} or relativistically boosted dark matter~(BDM)~\cite{Agashe:2014yua,Kong:2014mia,Berger:2014sqa,Alhazmi:2016qcs,Kim:2016zjx,Giudice:2017zke,Kim:2018veo,Chatterjee:2018mej}, that would induce signatures in detectors different from those of more traditional dark matter candidates. Because many of these new types of dark matter would produce unconventional signatures within a detector, few have been studied by typical dark matter search experiments.

One newly suggested model includes a relativistic dark matter particle that is boosted by annihilation of heavier dark matter particles in the Galactic Center or in the Sun~\cite{Agashe:2014yua,Kong:2014mia,Berger:2014sqa}. This would require at least two species of dark matter particles, denoted by $\chi_0$ and $\chi_1$ for the heavier and the lighter dark matter particle, respectively~\cite{Agashe:2014yua,Belanger:2011ww}. The first direct search for BDM was performed with the Super-Kamiokande detector by searching for energetic electron recoil signals induced by an elastic scattering of BDM~\cite{Kachulis:2017nci}. 

In addition to the elastic scattering of BDM, inelastic interactions in the recoil target are also possible. A few studies suggest interesting search channels where the scattered dark sector particle~(denoted as $\chi_2$) is different from the incoming $\chi_1$~\cite{Kim:2016zjx,Giudice:2017zke}. If $\chi_2$ is heavier than $\chi_1$, the subsequent decay of $\chi_2$ into a lighter state with visible standard model particles, such as an electron-positron pair, is possible~\cite{Kim:2016zjx,Giudice:2017zke,Chatterjee:2018mej}. This decay scheme is of  particular interest because it would enable new searches of BDM using many of existing detectors for the dark matter direct detections.

\begin{figure*}[!htb] 
\begin{center}
\includegraphics[width=0.9\textwidth]{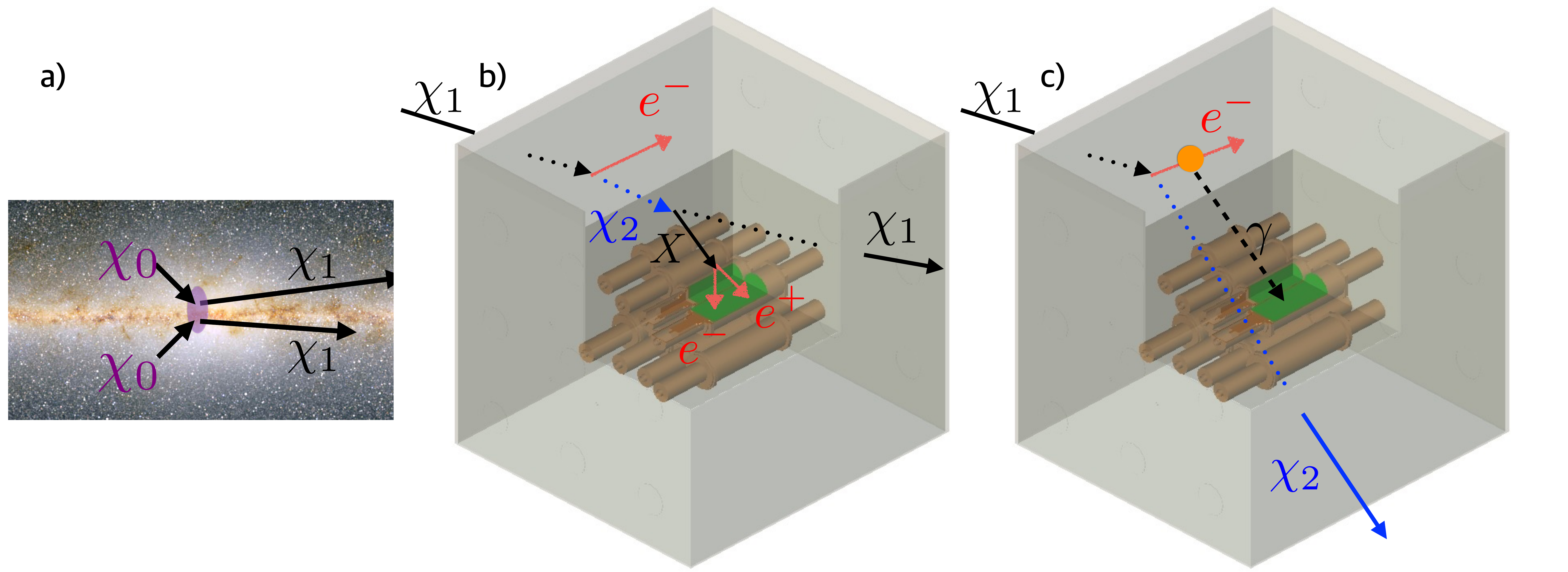} 
\caption{
  (a) Production of relativistic BDM $\chi_1$ in the Galactic Center by annihilation of a heavier dark matter $\chi_0$. (b) Illustration of multiple-site hits from an inelastic interaction of BDM for the case of two interactions occurring in two different NaI(Tl) and LS detectors.  (c) Illustration of Bremsstrahlung radiation-induced hits on two NaI(Tl) or LS detectors.
}
\label{fig_iBDM}
\end{center}
\end{figure*}

In this Letter, we report the first search for inelastic boosted dark matter~(iBDM) in a terrestrial detector, performed with the COSINE-100. 
The main purpose of the COSINE-100 experiment~\cite{Adhikari:2017esn,Adhikari:2017esn,Adhikari:2018ljm} is to confirm or refute the long-debated claim by the DAMA collaboration of dark matter discovery, which comes in the form of an annual modulation signal in event rates with energies between 1 and 6\,keV~\cite{Bernabei:2013xsa,Bernabei:2018yyw} corresponding to typical dark matter nucleon interactions~\cite{Savage:2008er}. Thus, in this Letter we also demonstrate the ability of detectors designed to search primarily for the low-energy nuclear recoils of the conventional dark matter interactions to detect high-energy electron recoils of the iBDM interactions.

In the Galactic Center, it is theorized that the boosted, lighter $\chi_1$ is produced by the pair-annihilation of two heavier $\chi_0$, as depicted in Fig.~\ref{fig_iBDM}~(a), with a total flux~\cite{Agashe:2014yua}
\begin{equation}
		\mathcal{F} = 1.6 \times 10^{-4} \mathrm{cm}^{-2} \mathrm{s}^{-1} \left(\frac{<\sigma v>_{0 \rightarrow 1}}{5 \times 10^{-26} \mathrm{cm}^{3} \mathrm{s}^{-1}} \right) \left(\frac{\mathrm{GeV}}{\mathrm{m}_0}\right)^{2},
\label{flux}
\end{equation} 
where the reference value for $<\sigma v>_{0 \rightarrow 1}$, or the velocity-averaged annihilation cross section of $\chi_0\chi_0 \rightarrow \chi_1 \chi_1$, corresponds to a correct dark matter thermal relic density for $\chi_0$ that is derived by a so-called ``assisted'' freeze-out mechanism~\cite{Belanger:2011ww} and m$_0$ denotes the mass of $\chi_0$. Production of BDM through the annihilation of $\chi_0$ is subject to uncertainties on the dark matter halo models~\cite{Navarro:1995iw,Kravtsov:1997dp,Moore:1999gc}. Here we assume the NFW halo profile~\cite{Navarro:1995iw} as described in Ref.~\cite{Agashe:2014yua}. 
The relativistic $\chi_1$ travels and interacts with terrestrial detector elements either elastically or inelastically.
We consider $\chi_1 e^-$ interactions from inelastic channel via a mediator $X$ exchange where the scattered dark sector particle, $\chi_2$, differs from the incoming $\chi_1$, i.e., $\chi_1 e^- \rightarrow \chi_2 e^-$. 
Furthermore, we require the mass of $\chi_2$~(m$_2$) to be heavier than the mass of $\chi_1$~(m$_1$) such that $\chi_2$ subsequently decays into $\chi_1$ with visible standard model particles such as $e^{-}e^{+}$ pairs, i.e., $\chi_2 \rightarrow \chi_1 e^{-}e^{+}$, mediated by $X$. Both on-shell (m$_X$ less than m$_2$-m$_1$ where m$_X$ denotes the mass of $X$) and off-shell (m$_X$ greater than m$_2$-m$_1$) decays of $\chi_2$ are possible~\cite{Giudice:2017zke}.

Considering the associated decay width of the three-body decay, the $\chi_2$ decay length can vary an order of 10~cm, which is sensitive range from the COSINE-100 detector, depending on m$_{X}$, the mass of $\chi_0$~(m$_0$), m$_1$, m$_2$, and the mixing parameter between the standard model particle and the dark sector particle~($\epsilon$). 
Following the calculation in Ref.~\cite{Kim:2016zjx,Giudice:2017zke}, an iBDM interaction would generate primary and secondary electrons and a positron with an energy within the range 1\,MeV--500\,MeV in the case of m$_0$=1\,GeV/c$^2$ as shown in Fig.~\ref{fig_iBDM}~(b).
Also, it is highly likely that energetic $e^-$ or $e^+$ produces a number of Bremsstrahlung radiations that deposit energies on the order of 1\,MeV as one can see in Fig.~\ref{fig_iBDM}~(c). For COSINE-100, we use the specific layout of the detector components combined with the predicted topology of iBDM interactions to identify candidate iBDM events.

\begin{figure}[!htb]
\begin{center}
\includegraphics[width=0.45\textwidth]{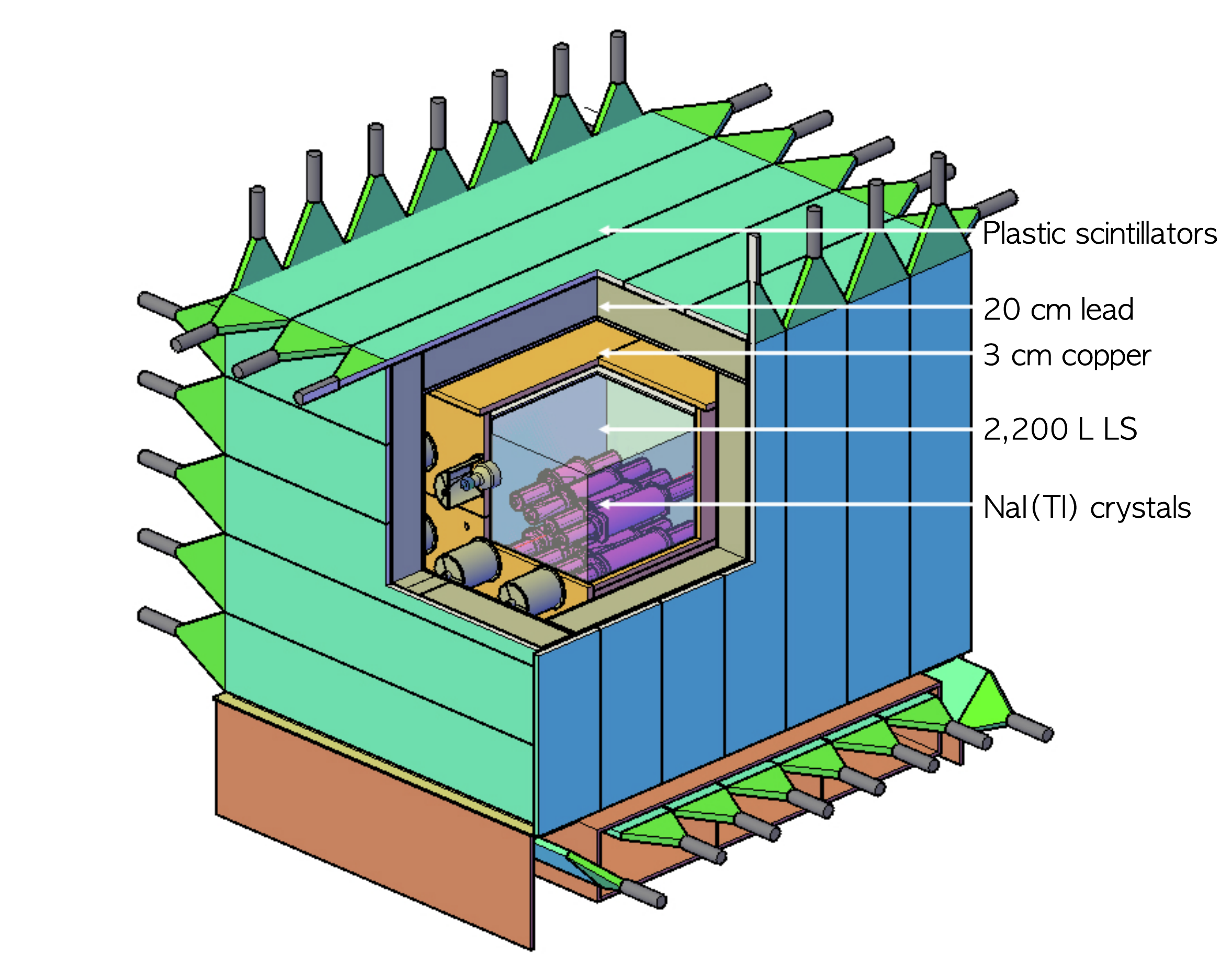} 
\caption{
		Schematic of the COSINE-100 detector. The NaI(Tl)~(106~kg) detectors are immersed in the 2,200\,L LAB-LS that are surrounded by layers of shields.
}
\label{fig_det}
\end{center}
\end{figure}

COSINE-100~\cite{Adhikari:2017esn} consists of a 106\,kg array of eight ultra-pure NaI(Tl) crystals each coupled to two photomultiplier tubes~(PMTs). The crystals are immersed in an active veto detector composed of 2,200\,L of linear alkylbenzene~(LAB)-based liquid scintillator~(LS)~\cite{Park:2017jvs}. The LS is contained within a 3\,cm thick shield of oxygen-free copper, a 20\,cm thick shield of lead, and an array of plastic scintillation counters for muon tagging~\cite{Prihtiadi:2017inr} as shown in Fig.~\ref{fig_det}. Crucial to this analysis is that the 2,200\,L of LS can act as an active detection volume that effectively creates a ton-scale detector for iBDM interactions.  Data obtained between 20 October 2016 and 19 December 2016 are used for this search with a total exposure of 59.5 live days. 

An event is triggered when coincident single photoelectrons in both PMTs coupled to a single crystal are observed within a 200\,ns time window. If at least one crystal satisfies the trigger condition, data from all other crystals and the LAB-LS are recorded. The LAB-LS signals do not generate triggers, except in the case of energetic muon events that are coincident with one of the muon detector panels~\cite{Adhikari:2018fpo}. 
Energy scales for the high energy region in the NaI(Tl) crystals and LS are calibrated with the 1465\,keV $\gamma$ line and the 2614\,keV $\gamma$ line from $^{40}$K and $^{208}$Tl, respectively. 
Geant4~\cite{Agostinelli:2002hh}-based simulations are used to understand the contribution of each background component~\cite{Adhikari:2017gbj,cosinebg}, as well as to verify energy scales and resolutions.

Event selections are based on the topology of iBDM events.
At first, we require the deposited energy in the LS be greater than 4\,MeV. 
We then remove muon-induced events that are tagged by the muon detector~\cite{Prihtiadi:2017inr}.
We require coincident hits in the NaI(Tl) crystals with the sum of deposited energies in all crystals be greater than 4\,MeV.
Finally, we reject $\alpha$-induced events in the crystals using a pulse shape discrimination method~\cite{kwkim15,adhikari16}. 
To summarize, the criteria for iBDM candidate events are:
\begin{description}
 \item[1] Energy of LS $>$ 4\,MeV
 \item[2] No selected muons from the muon detector
 \item[3] Total energy of the NaI(Tl) crystals $>$ 4\,MeV
 \item[4] No $\alpha$ events in the NaI(Tl) crystals
\end{description}

After applying these event selection criteria, we observe 21 candidate events from the 59.5\,days of the COSINE-100 data. Our study shows that a dominant background contribution is due to muons that pass directly through one side of the muon detector and stop in the LS or crystals. 
The muon detector's coverage is almost 4$\pi$, as there is a total of 37 panels attached to all six sides of the detector. 
The panels are spaced as closely as possible; however, small gaps between panels are inevitable because of the thickness of the materials used to wrap the panels~\cite{Prihtiadi:2017inr}. 
In the muon event selection, we tagged events as muon candidate events when a coincident signal between at least one of the muon panels and the LS is detected. 
However, if a muon passes through a gap between panels and stops in the LS or crystals, it cannot be tagged. 

To understand the missed-tag rate of the muon detector, we studied one specific type of muon candidate event that is tagged by a coincident signal between the bottom-side muon panels and the LS. 
Such a signal could be induced by an upward-going muon. However, the rate of upward-going muons is extremely small~\cite{Fukuda:1999pp} and most of the selected candidate events are actually downward-going muons that have passed through another side of the muon detector untagged. 
We estimate the untagged muon ratio~(r$_{\mathrm{untag}}$) as the following equation,
\begin{equation}
		\mathrm{r}_{\mathrm{untag}}=\frac{N_{\mathrm{bottom-LS}}}{N_{\mathrm{bottom-ALL}}},
\end{equation}
where $N_{\mathrm{bottom-LS}}$ is the number of tagged muons selected by only the bottom-side panels and LS coincidence, and $N_{\mathrm{bottom-ALL}}$ is all tagged muons with hits in the bottom-side panels. 
We obtain r$_{\mathrm{untag}}$=2.14$\pm$0.21\%. 
To determine the expected background rate from muon events, we select a sideband of muon-tagged candidate-like events. These events undergo the same selection criteria on the LS and NaI(Tl) crystals used for the iBDM candidate selection, and an additional muon tagging by the muon detector~\cite{Prihtiadi:2017inr} are required. The total number of these muon-tagged sideband events is multiplied by r$_{\mathrm{untag}}$\footnote{Depending on the number of muon sides~(N) satisfying the muon selection requirement, we multiply the untagging ratio N times.} which gives an expected background of
16.4$\pm$2.1 events. This is consistent with the 21 candidate events. 

\begin{figure}[!htb]
\begin{center}
\includegraphics[width=0.45\textwidth]{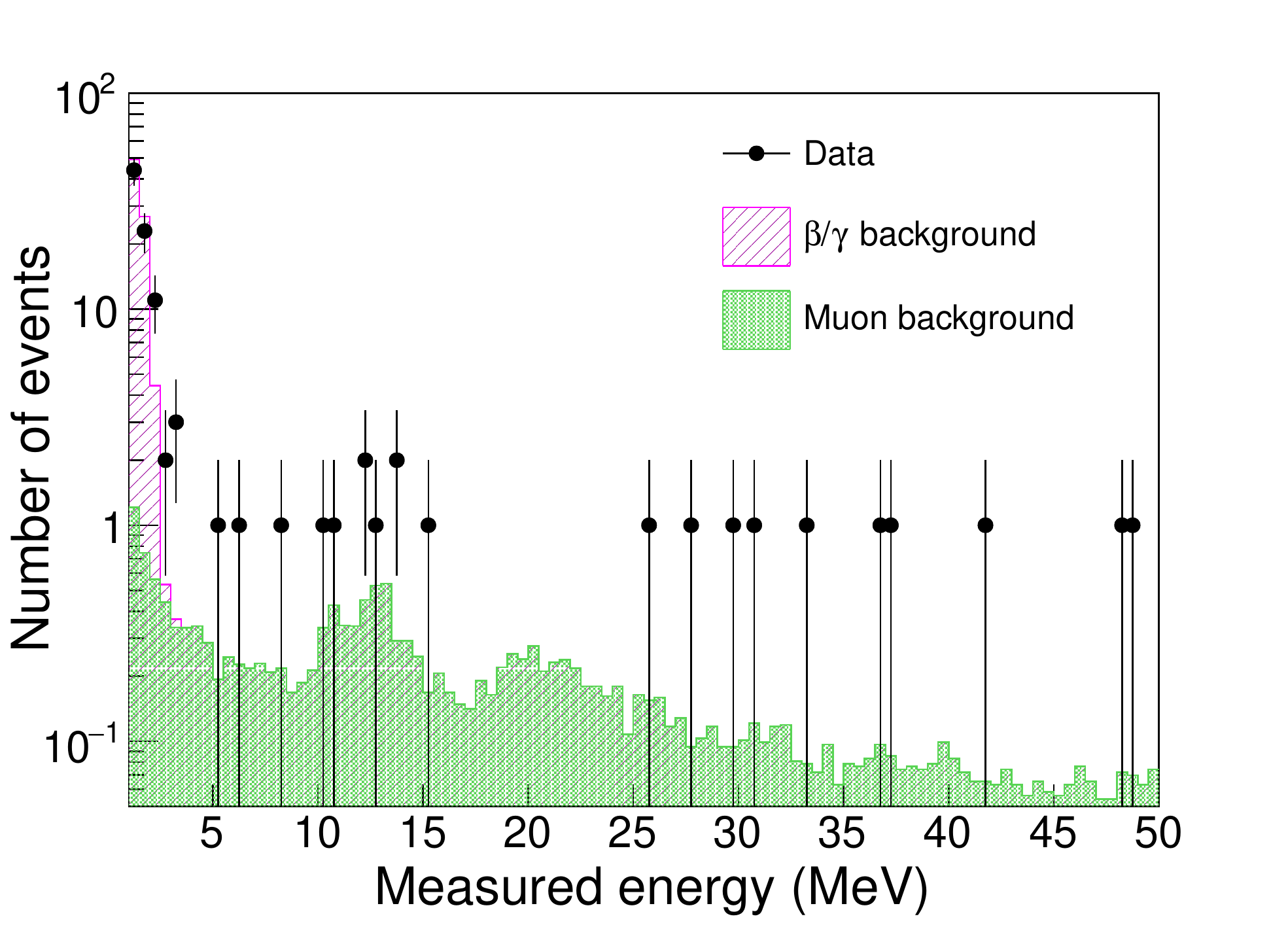} 
\caption{
  Spectrum of the summed energy from all crystals~(filled circle), after application of all selection criteria of the iBDM candidate events but with no requirement on the amount of energy observed in the crystals, is
	compared with the expected background (solid histogram). Contributions to the background from muon and $\beta$/$\gamma$ events caused by radionuclide contaminations are indicated. In the region of interest (energy greater than 4~MeV), muons are the only contribution. 
}
\label{fig_bg}
\end{center}
\end{figure}

Figure~\ref{fig_bg} shows the summed energy from all NaI(Tl) crystals after application of all selection criteria but with no requirement on the amount of energy observed in the crystals. 
Below 3\,MeV, the main background contribution is from $\beta$ and $\gamma$ rays originating from the detector and the surrounding materials. This background has been previously studied and is well understood~\cite{cosinebg}.
We model the background of the muons by using the muon-tagged sideband events that are reweighted with r$_{\mathrm{untag}}$ using the physics data.  
As one can see in Fig.~\ref{fig_bg}, our data are consistent with the known background contributions within statistical uncertainty.
In the region of interest (ROI), the muon-induced background is likely the only contribution.

\begin{figure}[!htb]
\begin{center}
\includegraphics[width=0.45\textwidth]{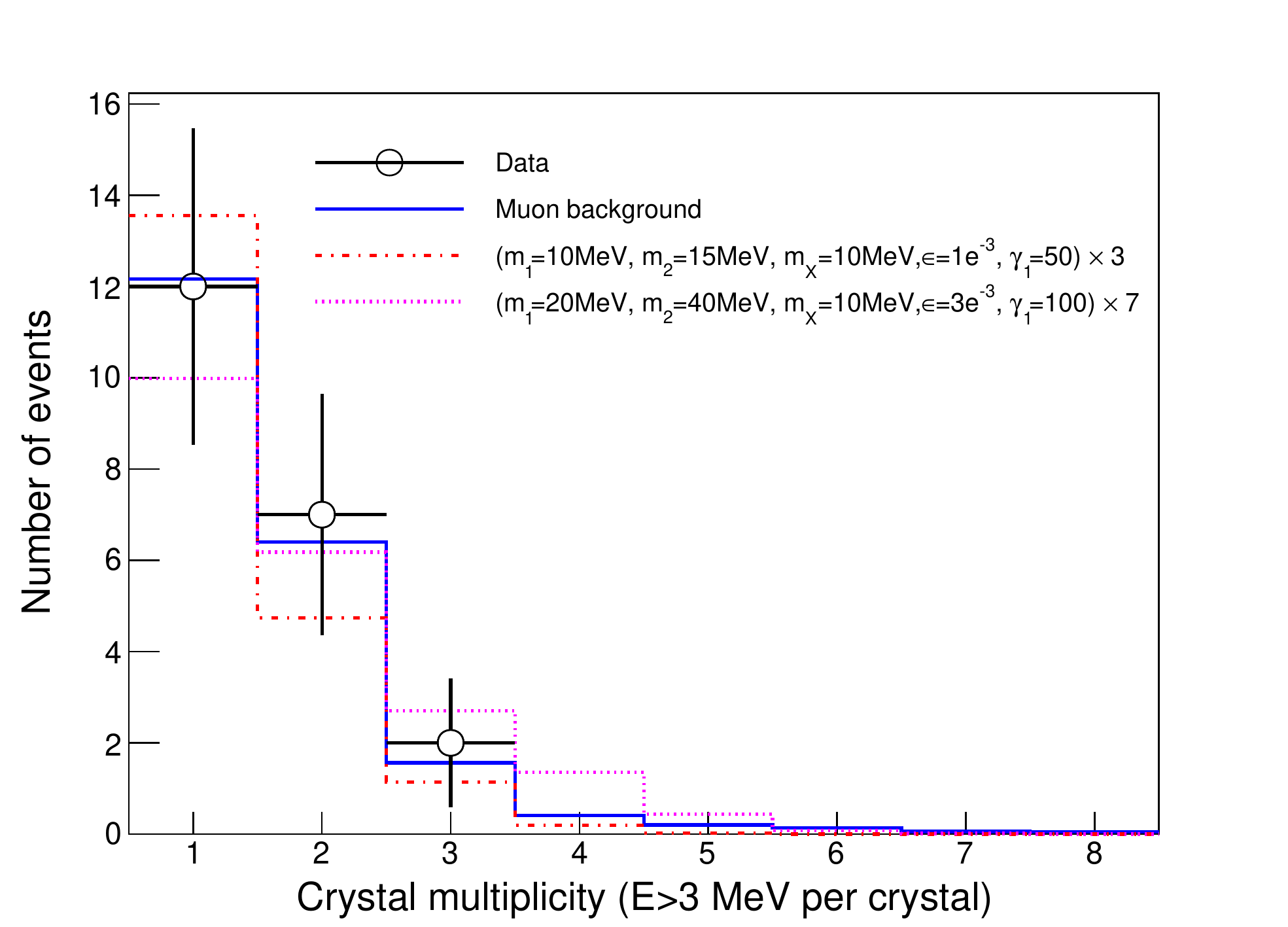} 
\caption{
Comparison of the event rate as a function of the crystal multiplicity in data with energy greater than 3\,MeV per crystal (open circle) and the expected background from muons (solid line).  	
	The expected signal shapes for two different model parameters~(dotted line and dashed line) are also shown. For easy comparison of the shapes, the dotted line and dashed line spectra are multiplied by a factor 7 and 3, respectively. 
}
\label{fig_signals}
\end{center}
\end{figure}

Because of this background contamination in the ROI, we examine parameters that demonstrate discrimination power between the muon background and iBDM induced signal.
A limited dynamic range beyond 5\,MeV for each crystal makes the summed energy of the crystals difficult to be used as the parameter.
Instead of the summed energy, we employ a crystal multiplicity selection that counts the number of crystals with energy depositions greater than 3\,MeV per crystal. 
For comparison of various model parameters, we generate iBDM signals following the calculations in Ref.~\cite{Kim:2016zjx,Giudice:2017zke} with the same model setup. Here, we assume $<\sigma v>_{0 \rightarrow 1} = 5\times10^{-26}$cm$^3$s$^{-1}$ in Eq.~\ref{flux} and fermionic dark matter. We consider a few choices of the iBDM model parameters~($\gamma_1$=m$_0$/m$_1$, m$_2$, m$_X$, and $\epsilon$). These events are then processed through the COSINE-100 detector simulation and the output events are subjected to the same selection criteria that are applied to the data. 
In Fig.~\ref{fig_signals}, the event rate of the data as a function of crystal multiplicity  is compared with that of the muon background as well as the theorized event rate from iBDM-induced signals for two different sets of model parameters. As one can see in the plot, the data are in good agreement with the muon background while the iBDM-induced signals can show a noticeable shape difference depending on the model parameters, especially with the kinetic energies of the charged electrons and positron in the primary and secondary interactions.

\begin{figure}[!htb]
\begin{center}
\includegraphics[width=0.45\textwidth]{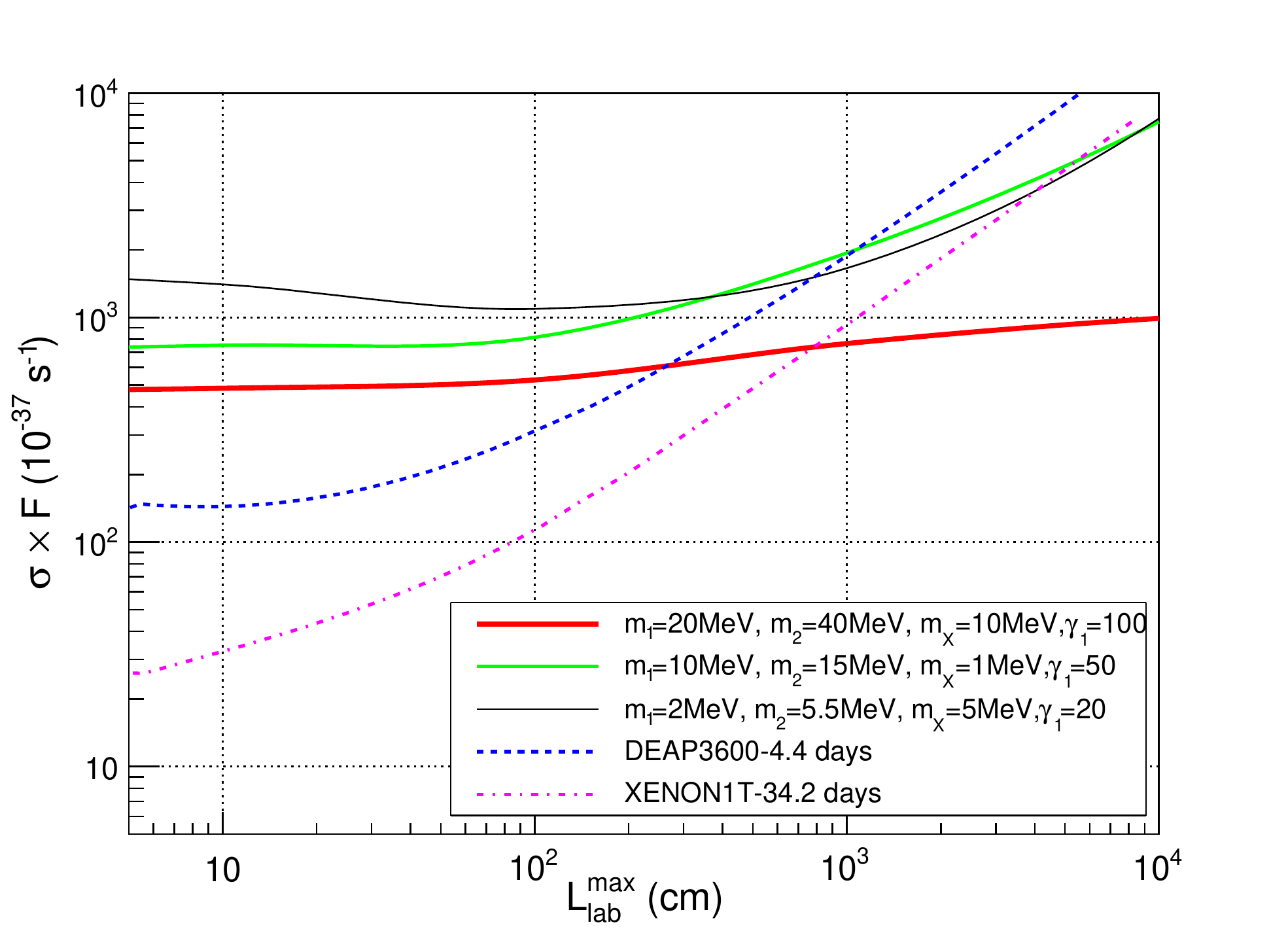} 
\caption{
  Measured 90\% CL upper limits from 59.5\,days of COSINE-100 data in the \lmax-$\sigma$ plane are presented for three different benchmark models. These results are compared with the experimental sensitivities of  XENON1T with 34.2\,days data~\cite{Aprile:2017iyp} and DEAP-3600 with 4.4\,days data~\cite{Amaudruz:2017ekt} calculated in Ref.~\cite{Giudice:2017zke}. 
}
\label{result_lmax}
\end{center}
\end{figure}

We express our results in the plane of \lmax-$\sigma$, where \lmax~ is the maximum mean decay length of the longest-lived particle, either $\chi_2$ or $X$, and $\sigma$ is the cross-section of the primary $\chi_1 e$ interaction. This is introduced in Ref.~\cite{Giudice:2017zke} for model independent comparison between existing and planned experiments of liquid Xe or Ar. In this case, they assume no background events if two tracks were produced in the fiducial volume with \lmax~ greater than the position resolution of each experiment.  We consider three different cases to represent our results where we have fixed the parameters of all dark sector particles in each case. Because \lmax~ depends on $(1/\epsilon)^2$, five different \lmax~ points~(5\,cm, 10\,cm, 100\,cm, 1,000\,cm, 10,000\,cm) have been scanned by simulating events with various values of $\epsilon$. These events are also processed through the detector simulation and the event selection. Selection efficiencies of the iBDM signals, which have primary interactions in the LS or NaI(Tl) crystals, vary between 1\% and 10\%, highly depending on the kinetic energy of $\chi_1$ and \lmax. 
To search for the iBDM-induced events, binned maximum likelihood fits to the measured crystal multiplicity are performed for the assumed signals. The Bayesian Analysis Toolkit~\cite{BAT} is used with probability density functions that are based on shapes of the simulated iBDM signal spectra and the muon background evaluated by the tagged-muon sideband events. Uniform priors are used for both the signals and the background. 

Data fits are performed for each of the signal models and we find no excess of events that could be induced by iBDM interactions. The posterior probabilities of signal are consistent with zero in all cases and 90\% confidence level~(CL) limits are obtained. Figure~\ref{result_lmax} shows the resulting limits from three different choices of the dark matter mass parameters. They are compared with the experimental sensitivities of XENON1T~(34.2\,days)~\cite{Aprile:2017iyp} and DEAP-3600~(4.4\,days)~\cite{Amaudruz:2017ekt} estimated with zero background assumption~\cite{Giudice:2017zke}. 
Benefiting from the large effective volume of the LS, measured limits from the COSINE-100 data are comparable with those of the ton-scale dark matter detectors' sensitivities. In particular, stronger limits at larger \lmax~ and heavier $\chi_0$~(corresponding to a larger boosting of $\chi_1$) are obtained. With a larger \lmax, the secondary decay may not happen in the detector volume. However, the modular crystal configuration of COSINE-100 is advantageous in terms of recording more multiple-site hits
from Bremsstrahlung radiation. 
The rate of Bremsstrahlung radiation is enhanced by the higher energy charged particles that occur with a larger $\gamma_1$.

\begin{figure}[!htb]
\begin{center}
\includegraphics[width=0.45\textwidth]{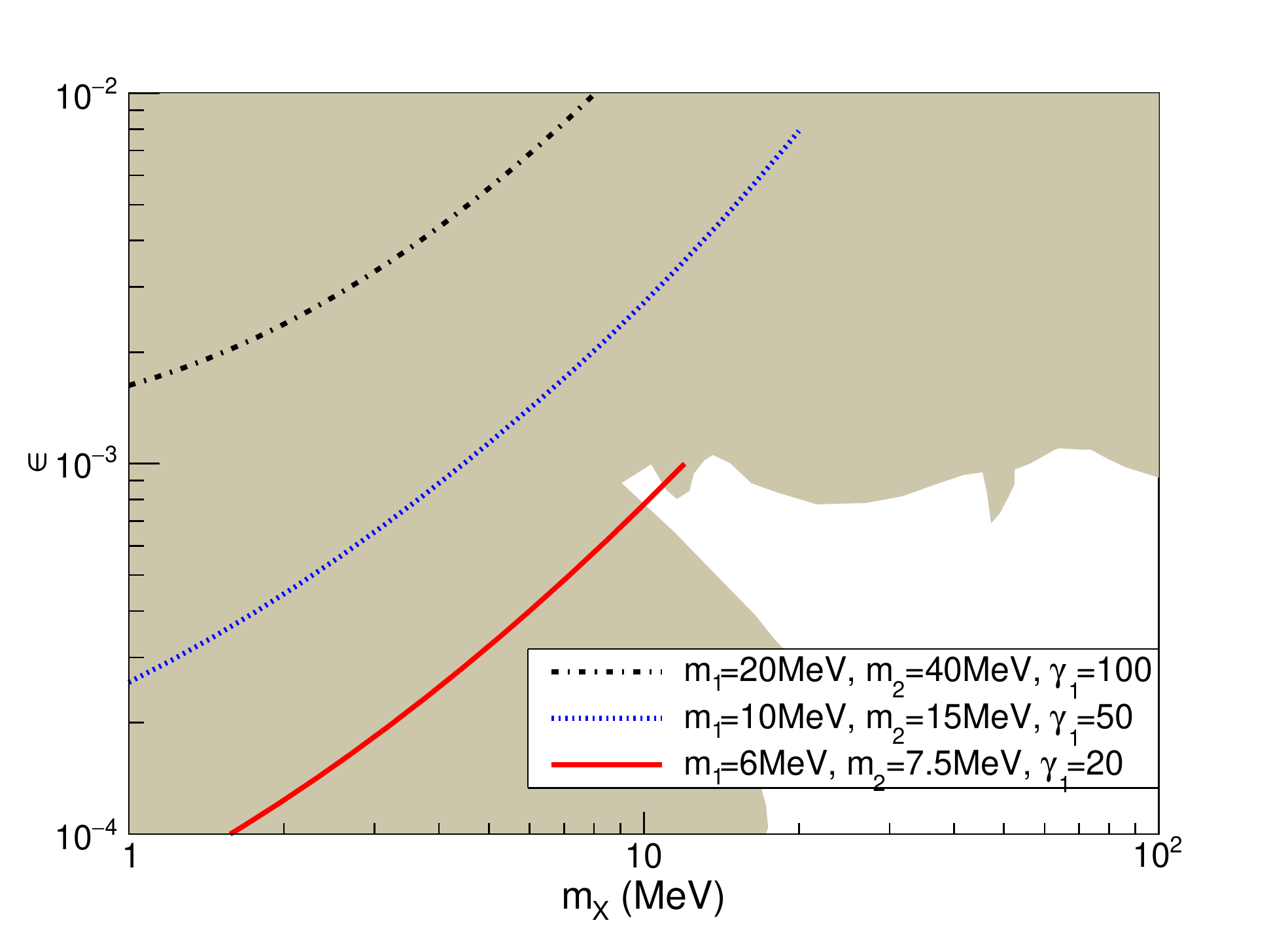} 
\caption{Measured 90\% CL upper limits~(lines) from 59.5\,days of COSINE-100 data in terms of the m$_X$ and $\epsilon$ parameters assuming the mediator to be a dark photon are presented for three different benchmark models. These results are compared with the currently excluded parameter space~(shaded region) from direct dark photon search experiments E141~\cite{Riordan:1987aw}, NA48~\cite{Batley:2015lha}, NA64~\cite{Banerjee:2018vgk}, Babar~\cite{Lees:2014xha}, and bounds from the electron anomalous magnetic moment~($g$-2)$_{e}$~\cite{Davoudiasl:2014kua}. 
}
\label{result_dp}
\end{center}
\end{figure}

It is possible to interpret this result in the context of the dark photon phenomenology by assuming that the interaction between the standard model particles and the dark sector particles is mediated by a dark photon. This is interesting because it allows results obtained from this analysis to be compared with other dark photon searches, which are typically expressed in terms of  the parameters m$_X$ and $\epsilon$. In our analysis, we generate signals from three different sets of model parameters by fixing m$_1$, m$_2$, and $\gamma_1$ while varying m$_X$. In this event generation, we assume fully inelastic scattering. Because of the invisible decay of $X$, we only consider m$_X <2$m$_1$. 
Figure~\ref{result_dp} shows the measured 90\% CL upper limits from the COSINE-100 data for the aforementioned model parameters compared with existing constraints from dark photon search experiments. 
In the specific case of m$_1$=6\,MeV, m$_2$=7.5\,MeV, and $\gamma_1$=20, our limits begin to explore parameter regions that have not been searched by any other experiment, though the specific model discussed in this paper needs to be assumed.

In summary, we have performed a first direct measurement for evidence of inelastic boosted dark matter by searching for energetic electrons and positrons produced in the COSINE-100 detector with 59.5\,days of data. The COSINE-100 detector has a unique advantage in detecting this signal because of the 2,200\,L of LAB-LS that surrounds 106.5\,kg of low-background NaI(Tl) crystals. No signal excess is found and, therefore, 90\%\,CL limits are set for various model parameters. An interpretation of a dark photon interaction further explores a new parameter space that has not been tested by other direct dark photon search experiments, though it is highly model dependent. The COSINE-100 detector has stable operation taking more than 2~years of good quality data. An updated study with a larger data sample and improved analysis techniques will allow us to search a much larger region of the currently unexplored parameter space.

\acknowledgments
We thank Jong-Chul Park for encouraging this analysis and for insightful discussions. 
We also acknowledge  Seodong Shin for insightful discussions.
We thank the Korea Hydro and Nuclear Power (KHNP) Company for providing underground laboratory space at Yangyang.
This work is supported by:  the Institute for Basic Science (IBS) under project code IBS-R016-A1 and NRF-2016R1A2B3008343, Republic of Korea;
UIUC campus research board, the Alfred P. Sloan Foundation Fellowship,
NSF Grants No. PHY-1151795, PHY-1457995, DGE-1122492,
WIPAC, the Wisconsin Alumni Research Foundation, United States; 
STFC Grant ST/N000277/1 and ST/K001337/1, United Kingdom;
and Grant No. 2017/02952-0 FAPESP, CAPES Finance Code 001, Brazil.
\end{document}